\documentclass[%
 reprint,
 amsmath,amssymb,
 aps,
floatfix,
fleqn,
]{revtex4-1}

\usepackage{graphicx}% Include figure files
\usepackage{dcolumn}% Align table columns on decimal point
\usepackage{bm}% bold math
\usepackage{comment}
\usepackage{color}
\newcommand{\REV}[1]{{\color{magenta} #1}}
\newcommand{\TBR}[1]{{\color{blue} #1}}

\begin{document}

\preprint{\TBR{APS/123-QED}}

\title{Collective dynamics of active filament complexes
                                        }% Force line breaks with \\
%\thanks{A footnote to the article title}%

\author{Hironobu Nogucci}
%\altaffiliation[Also at ]{Physics Department, XYZ University.}%Lines break automatically or can be forced with \\
%author{S. Ishihara  }%
 \email{noguchi@complex.c.u-tokyo.ac.jp}
\affiliation{%
 Department of Basic Science, Graduate School of Arts and Sciences, The University of Tokyo, 3-8-1 Komaba, Meguro, Tokyo, Japan
%This line break forced with \textbackslash\textbackslash
}%

%\collaboration{MUSO Collaboration}%\noaffiliation

\author{Shuji Ishihara}
%\homepage{http://www.Second.institution.edu/~Charlie.Author}
\affiliation{
 Department of Physics, School of Science and Technology, Meiji University, 1-1-1 Higashi-Mita, Tama, Kawasaki, Kanagawa, Japan\\
 % This line break forced% with \\
}%
%\affiliation{
% Third institution, the second for Charlie Author
%}%
%\author{Delta Author}
%\affiliation{%
% Authors' institution and/or address\\
% This line break forced with \textbackslash\textbackslash
%}%

%\collaboration{CLEO Collaboration}%\noaffiliation

\date{\today}% It is always \today, today,
             %  but any date may be explicitly specified

\begin{abstract}
%An article usually includes an abstract, a concise summary of the work
%covered at length in the main body of the article.

Networks of biofilaments are essential for the formation of cellular structures that support various biological functions.
For the most part, previous studies have investigated the collective dynamics of rod-like biofilaments; however, the shapes of the actual subcellular components are often more elaborate.
In this study, we considered an active object composed of two active filaments, which represents the progression from rod-like biofilaments to complex-shaped biofilaments.
Specifically, we numerically assessed the collective behaviors of these active objects in two dimensions and observed several types of dynamics depending on the density and the angle of the two filaments as shape parameters of the object.
Among the observed collective dynamics, a moving density band that we named a `moving smectic' is introduced here for the first time.
By analyzing the trajectories of individual objects and the interactions among them, this study demonstrated how interactions among active biofilaments with complex shapes could produce collective dynamics in a non-trivial manner.

\begin{description}
%\iffalse
%\item[Usage]
%Secondary publications and information retrieval purposes.
\item[PACS numbers]
87.16.Ka
05.65.+b
%May be entered using the \verb+\pacs{#1}+ command.
%\item[Structure]
%You may use the \texttt{description} environment to structure your abstract;
%use the optional argument of the \verb+\item+ command to give the category of each item.
%\fi
\end{description}
\end{abstract}
\pacs{Valid PACS appear here}% PACS, the Physics and Astronomy
                             % Classification Scheme.
%\keywords{Suggested keywords}%Use showkeys class option if keyword
                              %display desired
\maketitle
%\tableofcontents
%\iffalse
%\section{\label{sec:level1}First-level heading:\protect\\ The line
%break was forced \lowercase{via} \textbackslash\textbackslash}
%\fi

\section{\label{sec:Intro}Introduction}
Organized networks of biofilaments such as microtubules and actin fibers play essential roles in the formation, maintenance, and alteration of cell structures \cite{howard2001mechanics}.
For example, cortical microtubules in plant cells exhibit aligned structures called `bundles', which mechanically support the cells \cite{Dixit2004a,Baulin2007,Hawkins2010, Hawkins2010a}.
These subcellular structures composed of biofilaments are usually formed through active processes driven by energy use in hydrolysis.
Polymerization and depolymerization cause length changes to the microtubules and cause them to collide with each other, which ultimately leads to their global alignment.
Another active mechanism involves interactions between biofilaments and molecular motors.
When the molecular motors bind to the filaments and march along them,
active forces act on the filaments and drive the formation of spatio-temporal filament patterns \cite{Straub1950,Gibbons1963,Janmey1990,Grieder2000,Sanchez2011}.
To date, a number of {\it in vitro} experiments have revealed various types of structures resulting from this motor activity \cite{Karsenti2006,Moore2009,Schaller2010}.
For example, filaments form locally ordered patterns such as ray-like asters and vortices, depending on the type and concentration of molecular motors present \cite{Ndlec1997,Surrey2001,Liu2011,Sumino2012}.
In addition, the collective motion of biofilaments can emerge via their active interaction with each other, which has led to the interesting paradigm of `active matter' \cite{Vicsek1995,Gregoire2004,Chate2006,Ramaswamy2010, Peruani2011, Marchetti2013}.

Many theoretical studies have attempted to address the self-organization of active filaments. Based on the contracting force of active filaments driven by molecular motors, coarse-grained and continuum kinetic equations were proposed to explain the inhomogeneous accumulation of filaments (i.e., bundles) \cite{Kruse2000,Kruse2003} and local structures such as vortices and asters \cite{Kruse2005}.
Similar global patterns were observed in another continuum model in which nematic collisions were taken into account
 \cite{Liverpool2003,Liverpool2005,Ahmadi2005,Liverpool2006,Ahmadi2006}.
To explain the emergence of vortices and asters from microscopic processes,
Aranson and Tsimring introduced simple stochastic rules for inelastic collisions of biofilaments, and thus
derived continuum equations by which transport coefficients could be related to microscopic physical quantities \cite{Aranson2005,Aranson2006}.
Aranson's group performed Monte Carlo simulations of the elementary alignment processes to verify their theory \cite{Jia2008}; however, a correspondence with continuum theory remains elusive.
In other studies simulating the dynamics of filaments, new patterns, including stripes, were found \cite{Head2013}, and
the viscoelasticity of the networks of biofilaments was discussed \cite{Broedersz2014}.

The majority of previous studies mainly considered objects with rod-like shapes as the simplest biofilaments. However,{\it in vivo}, filaments often bind to each other with additional related molecules and form molecular complexes, which constitute a functional unit in various biological processes.
For instance, microtubules on mitotic spindles elongate radially from a pair of centrosomes, which separate a mother cell into two daughter cells \cite{Campas2008}. Beneath the apical membrane of a multi-ciliated cell, the root of the cilium, i.e., the basal body, has an appendage known as the basal foot that functions as a microtubule organizing center.
Microtubules are generated from these basal feet and subsequently connect with each other to become organized as a cell-sized network, which is suggested to be involved in the ordered alignment and direction of the beating cilia \cite{Kunimoto2012,Werner2011}.
In this example, microtubules that are pivoted by the basal body can act as a functional unit, and their interaction may lead to alignment of the cilia.
However, the mechanisms underlying the function of such biofilament complexes, especially the relationship between the shape of the molecular complexes and their emerging dynamics, remain largely unknown.
In addition, current developments in nanobioengineering, such as optical manipulation \cite{Dinu2009}, have enabled the design of biofilament complexes that can give rise to new types of the self-organization of biofilaments.
Such technological progress has also motivated us to explore the possible dynamics of biofilaments when they do not have a simple rod-like shape.

To gain fundamental insights into these mechanisms, in the present study, we performed numerical simulations of molecular complexes composed of two filaments, named `active filament complexes (AFCs)'. This model represents a simple investigative progression from rod-like filaments to complex-shaped filaments.
In our model, the active interactions of filaments via molecular motors are considered
in two dimensions, whereas collisions and excluded volume interactions are ignored.
We discovered interesting dynamics depending on the density and shape of AFCs, including a newly identified dynamic form termed a `moving smectic' (described in detail later) that has not been previously reported.
By tracking trajectories and investigating AFC interactions, we also explored the mechanisms by which the various dynamic patterns observed might arise in the system.

The remaining sections of this paper are organized as follows.
The detailed model is described in Sec.~\ref{sec:model}.
In Sec.~\ref{sec:pattern}, the observed dynamics in numerical simulations are classified according to quantities such as ferromagnetic (polar) and nematic (apolar) order parameters.
By tracking individual AFCs and characterizing the crossing of filaments, analyses are described from Sec.~\ref{sec:motion} to Sec.~\ref{sec:cross}.
%%The dynamics of the modified model system are briefly discussed in Sec.~\ref{sec:heavy}.
Finally, a comparison with earlier studies and the potential for future experiments are discussed in Sec.~\ref{sec:disc}.

%%%%%%%% Model Part %%%%%%%%%%%%%%
\section{\label{sec:model}Model}

\subsection{\label{subsec:model1}Dynamics of single active filaments}
\begin{figure}[b]
\includegraphics[width=8.6cm,clip]{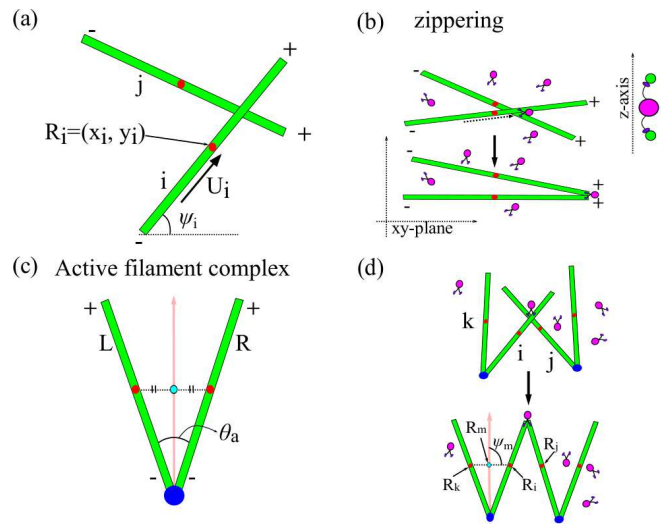}
\caption{\label{fig:zip}
Schematic representations of the interaction between filaments and active filament complexes (AFCs).
(a) Configuration of crossing single filaments.
Red points represent the center of filaments.
(b) Possible alignment of crossed filaments (`zippering').
(c) AFC with shape angle $\theta_a$ and length $\ell ~\rm{nm}$.
Each filament is named as an L- or R-~filament.
The center of mass in the AFC is located on the bisecting arrow (red).
The cyan point denotes the center of mass of an AFC in the light-anchor model.
(d) Interaction and possible alignment of crossed AFCs.
}
\end{figure}

To model the dynamics of single active filaments, the formulation given by Tanase~\cite{Tanase2004} was modified.
In this formulation, a biofilament is described as a single rod diffusing in a viscous liquid.
Each filament possesses polarity associated with its elongating direction from the minus to plus end.
Therefore, $3N$ variables are sufficient to represent the configuration of the microtubules, i.e., their center position, ${\bf R}_i=(x_i,y_i)$, and the direction of their plus end, $\psi_i$ ($\in[0,2\pi]$) [Fig.~\ref{fig:zip}(a)].

In a viscous environment where hydrodynamic interactions are negligible, the dynamics of a rod-like filament are governed by $\dot{\bf R}_i = \hat{{\boldsymbol \zeta}}^{-1} {\bf F}_i$ and $\dot{\theta}_i = \zeta^{-1}_{rot}(xF_{iy} - yF_{ix})$. Here, ${\bf F}_i = (F_{ix}, F_{iy})$ is the external point force acting on the filament, and ${\bf x} = (x, y)$ is the relative vector pointing from the center of mass of the filament to the point of action.
$\hat{{\boldsymbol \zeta}}^{-1} = \zeta^{-1}_{\parallel} {\bf U}_i {\bf U}_i + \zeta^{-1}_{\perp} ({\bf I} -{\bf U}_i {\bf U}_i)$ represents the reciprocal matrix of orientation-dependent drag coefficients, where ${\bf U}_i=(\cos{\theta_i}, \sin{\theta_i})$ is the unit orientation vector parallel to the filament.
$\zeta_{\parallel}$, $\zeta_{\perp}$, and $\zeta_{rot}$ are drag coefficients associated with the longitudinal, transversal, and rotational motions of filaments, respectively.
These coefficients in dilute systems are related as $\zeta_{\parallel} = \zeta_{\perp}/2$, $\zeta_{rot} = \frac{\ell^2}{c}\zeta_{\parallel}$, where $\ell$ is the filament length and constant $c = 6$ is independent of $\ell$ at the limit of the high aspect ratio of the rods \cite{DoiEd1986,Tao2005}.
We fixed $\ell$ by ignoring polymerization/depolymerization and approximate $\zeta = \zeta_{\parallel} \simeq \zeta_{\perp}$, since their difference is only a factor of $2$ \cite{Aranson2006}.
Then, the dynamics of ${\bf R}_i$ has the simple form $\dot{\bf R}_i = \zeta^{-1}{\bf F}_i$.

The point force to the $i$-th filament, ${\bf F}_i$, derives from two independent parts.
One is the thermal noise triggered by the random motion of the solvent molecules.
The other is the motor-mediated interaction with other filaments, which is our main point of interest, formulated as follows.
In the cell cytoplasm or solvents, many molecular motors such as kinesin and dynein bind to the filaments.
While kinesin binds, a sliding motion occurs along the filament from the minus to the plus end.
Because most molecular motors bind to the filaments by multiple binding heads, when two filaments are in close proximity and are simultaneously bound to the same motors, the slidings of the motors result in mechanical torque that orientates the two filaments by generating tension $T_m$ at the binding point ${\bf R}_c$, a process known as `zippering' [Fig.~\ref{fig:zip}(b)] \cite{Vale2000,Dixit2004a}.
In this process, molecular motors diffuse $\sim \! \! 100$ times faster than biofilaments, and can hence be regarded as uniformly distributed \cite{Aranson2006}.
In addition, under the high concentration of molecular motors, the binding to filaments and sliding along them occur at a constant rate.
These conditions enable us to treat $T_m$ as a fixed parameter without explicitly considering motor dynamics.
We assume that the direction of the zippering force generated by the molecular motors only depends on the relative angles between a contacting pair of two filaments [Fig.~\ref{fig:zip}(a) and (b)]\cite{Tanase2004}.
Taken together, for a two-dimensional system in which crossings of
filaments are not prevented (see below), 
the equations of motion for the $i$-th filament are expressed as
\begin{eqnarray}
\dot{{\bf R}}_{i}&=& \sum_{j \in C_i} \Biggl[\frac{T_m}{\pi\zeta}({\bf U}_{j}-{\bf U}_{i}) \Biggr] + {\boldsymbol \eta}_{i}, \label{eq:z1}\\
\dot{\psi}_{i}&=& \sum_{j \in C_i} \Biggl[\frac{T_m}{\pi\zeta_{rot}}\bigl[ ({\bf R}_{c}-{\bf R}_{i})_x ({\bf U}_{j})_y \nonumber \\
&-&({\bf R}_{c}-{\bf R}_{i})_y ({\bf U}_{j})_x\bigr] \Biggr] + {\eta}_{i}  \nonumber\\
&=&\sum_{j \in C_i} \Biggl[\frac{T_m}{\pi\zeta_{rot}}\bigl[ ({\bf R}_{j}-{\bf R}_{i})_x ({\bf U}_{j})_y \nonumber \\
&-&({\bf R}_{j}-{\bf R}_{i})_y ({\bf U}_{j})_x \bigr] \Biggr] + {\eta}_{i}^{rot}. \label{eq:z2}
\end{eqnarray}
The first terms of the right-hand side of the two equations represent `zippering', and the second terms represent thermal noise.
$C_i$ represents a set of filaments that interact with the $i$-th filament via motors.
If none of the filaments interacts, the first terms vanish.
${\boldsymbol \eta}_i (t)$ is a state-dependent random force obeying
${\boldsymbol \eta}_i (t)= \eta_i^{\parallel} (t) {\bf U}_i + \eta_i^{\perp} (t) {\bf V}_i$,
where ${\bf V}_i =(-\sin{\theta_i}, \cos{\theta_i})$ is the unit vector perpendicular to the $i$-th filament.
$\eta_i^{\parallel}$, $\eta_i^{\perp}$, and $\eta_i^{rot}$ represent Gaussian white noise following the statistics:~$\langle \eta_i^{\parallel}(t) \rangle = \langle \eta_i^{\perp}(t) \rangle  = \langle \eta_i^{rot}(t) \rangle = 0$, $ \langle \eta_i^{\parallel}(t) \eta_i^{\parallel}(t') \rangle = 2D_{\parallel} \delta (t - t')$, $ \langle \eta_i^{\perp}(t) \eta_i^{\perp}(t') \rangle = 2D_{\perp} \delta (t - t')$, $ \langle \eta_i^{rot}(t) \eta_i^{rot}(t') \rangle = 2D_{rot} \delta (t - t')$.
$D_{\parallel}$, $D_{\perp}$, and $D_{rot}$ are longitudinal, transversal, and rotational diffusion constants, and are related to drag coefficients
by the Einstein relations as $D = D_{\parallel} = D_{\perp} = k_BT\zeta^{-1}$ and
$D_{rot}=k_BT\zeta^{-1}_{rot}$, respectively, where $k_B T$ should be interpreted as the effective temperature.
In the equations, total momentum is conserved if thermal noise is eliminated.

In the derivation of the equations above, we ignored the excluded volume effect for simplicity, although it may contribute to apolar alignment~\cite{Ahmadi2005,Ahmadi2006}.
This can be justified with two assumptions [Fig.~\ref{fig:zip}(b)].
One is that we set a pseudo two-dimensional system whose length along the $z$-axis is around $0.1-1.0 ~\rm{\mu m}$.
Thus, filaments cannot lean toward the $z$-direction due to their length ($> \! 1~\rm{\mu m}$ for a microtubule), and their crossings are not prevented by physical contacts along the $x\!-\!y$ plane because both the diameter of the filaments%, $d_{fil}$,
($\sim \! 26~\rm{nm}$ for a microtubule) and the size of the molecular motors%, $s_m$,
($\sim \! 80~\rm{nm}$ for kinesin) are sufficiently small.
The other assumption concerns the motor property.
If the density of the motors and their affinity to the filaments are both high,
the filaments are always associated with multiple motors that cross-link and drive the zippering interaction.
Thus, in our setup of two-dimensional simulations, motor-mediated tension always acts at the crossing point and
zippering will have a greater effect than collision.
This situation holds for a microtubule network beneath the apical membrane and is consistent with many experimental setups.

\subsection{\label{subsec:model2}Model for AFCs}
To study the dynamics of active filaments with shape, the model equations for active single filaments were extended as follows.
As a simple AFC, we consider an object to which two filaments are anchored on their minus ends at an angle of $\theta_a$ [Fig.~\ref{fig:zip}(c) blue point].
The lengths of the two filaments are set as $\ell$.
Hereafter, $\theta_a$ is referred to as the shape angle, and the left and right filaments are called L- and R-~filaments, respectively.
We assume that the mass of the anchoring object is negligibly small compared to that of the filaments,
and the motion of the two filaments dominates the dynamics of AFCs.
For an AFC composed of the $i$-th and $k$-th filaments, the center of mass of the AFC becomes ${\bf R}_m = ({\bf R}_i + {\bf R}_k)/2$.
When an AFC interacts with other AFCs, the equations of motion for the center of mass and the angle, $\dot{R}_m$ and $\dot{\psi}_m \equiv \dot{\psi}_i = \dot{\psi}_k$,
become [Fig.~\ref{fig:zip}(d)],
\begin{eqnarray}
\dot{{\bf R}}_{m} = \frac{1}{2}&\Biggl[&\sum_{j \in C_i} \biggl[\frac{T_m}{\pi\zeta}({\bf U}_{j}-{\bf U}_{i}) \biggr]+ {\boldsymbol \eta}_{i} \nonumber \\
&+& (k \leftrightarrow i) \Biggr], \label{eq:z3}\\
\dot{\psi}_{m} = \frac{1}{2}&\Biggl[&\sum_{j \in C_i} \biggl[ \frac{T_m}{\pi\zeta_{rot}}\Bigl[ \Bigl({\bf R}_{c}-{\bf R}_m\Bigr)_x({\bf U}_j-{\bf U}_{i})_y \nonumber \\
&-& \Bigl({\bf R}_{c}-{\bf R}_m\Bigr)_y ({\bf U}_j-{\bf U}_{i})_x \Bigr] \biggr] + {\eta}_{i}^{rot} \nonumber \\
&+& (k \leftrightarrow i)  \Biggr]\nonumber \\
= \frac{1}{2}&\Biggl[& \sum_{j \in C_i} \biggl[\frac{T_m}{\pi\zeta_{rot}}\Bigl[ ({\bf R}_j-{\bf R}_{i})_x({\bf U}_j)_y  \nonumber \\
&-& ({\bf R}_j-{\bf R}_{i})_y ({\bf U}_j)_x \nonumber \\
&+&  [({\bf R}_i-{\bf R}_{k})_x({\bf U}_j-{\bf U}_i)_y \nonumber \\
&-& ({\bf R}_i-{\bf R}_{k})_y ({\bf U}_j-{\bf U}_i)_x]/2 \Bigr] \biggr] + {\eta}_{i}^{rot} \nonumber \\
&+& (k \leftrightarrow i) \Biggr], \label{eq:z4}
\end{eqnarray}
where ($k \leftrightarrow i$) represents the counterpart term to the first half term of each equation, permuating the indices $i$, $k$, and $C_i$ into $k$, $i$, and $C_k$, respectively.
Here, $C_i$ and $C_k$ are sets of filaments that interact with the $i$-th and $k$-th filaments, respectively.

This model is referred to as the `light-anchor model'.
In general, when the mass of filaments is $m_f$ and that of the anchor is $m_a$, the center of mass of the AFC is positioned at $\frac{m_f \ell \cos{(\theta_a/2})}{m_a + 2m_f}$ distance from the anchor along the bisection of the two filaments [Fig.~\ref{fig:zip}(c) red line].
The light-anchor model is justified in the case of $m_a \ll 2m_f$.
If noise is absent, the model also conserves total momentum and is classified into an `active wet system' according to Ref.~\cite{Marchetti2013}.

Numerical simulations of AFCs were performed in two-dimensional systems with periodic boundary conditions.
The initial states of AFCs are set to obey spatially and orientationally uniform distributions. The system size is fixed at $L \times L = 4.0 \times 10^{2}~{\rm \mu m^2}$ unless mentioned otherwise.
To evaluate the effect of the AFCs' shape on the collective dynamics, the shape angle $\theta_a$ is controlled.
At $\theta_a = 0^\circ$, the AFC coincides with a polar filament, whereas at $\theta_a = 180^{\circ}$, the AFC is a bipolar filament with length $2\ell ~\rm{\mu m}$.
The density of the AFCs is also controlled by changing the number of AFCs, $N$.
The other parameters are fixed as follows: filament length, $\ell = 2.0~\rm{\mu m}$;
translational drag coefficient, $\zeta = 0.4 \times 10^{-6} ~\rm{kg\cdot s^{-1}}$;
effective temperature, $k_BT = 4.0 \times 10^{-21} ~\rm{kg\cdot m^{2}\cdot s^{-2}}$;
and tension of the molecular motors acting on the crossing filaments, $T_m = 2.4 \times 10^{-12} ~\rm{kg\cdot m\cdot s^{-2}}$.
These parameter values are in a physiologically plausible range and are comparable to those used in previous studies \cite{Tanase2004,Jia2008}. At these values, the noise is quite weak, and thus the dynamics of the systems are mostly dominated by active forces.
The simulation time is taken as $t = 1000~\rm{s}$, which is sufficiently long compared to the relaxation times of all order parameters ($<100~{\rm s}$) introduced in Sec.~\ref{sec:pattern}.

\begin{figure}[b]
\includegraphics[width=8.0cm]{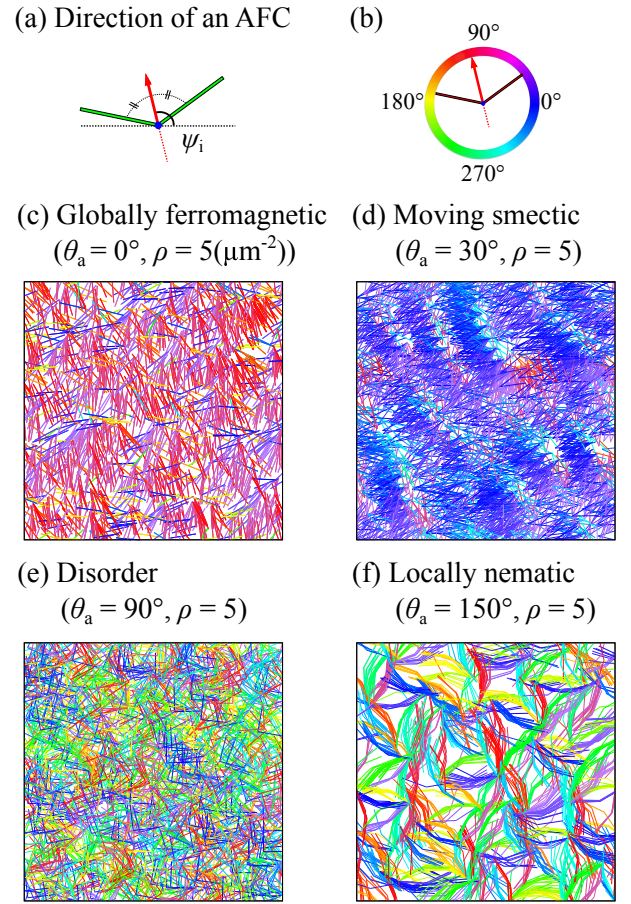}
\caption{\label{fig:snap1}
Snapshots of simulation results.
(a) Direction of an AFC, $\psi_{i}$, is defined as the direction bisecting two filaments for each AFC.
(b) Color code for the direction of AFCs.
(c)--(f) Snapshots of the simulation results at indicated parameter values. Filaments are colored according to their directions.
}
\end{figure}
\begin{figure}[b]
\includegraphics[width=8.6cm]{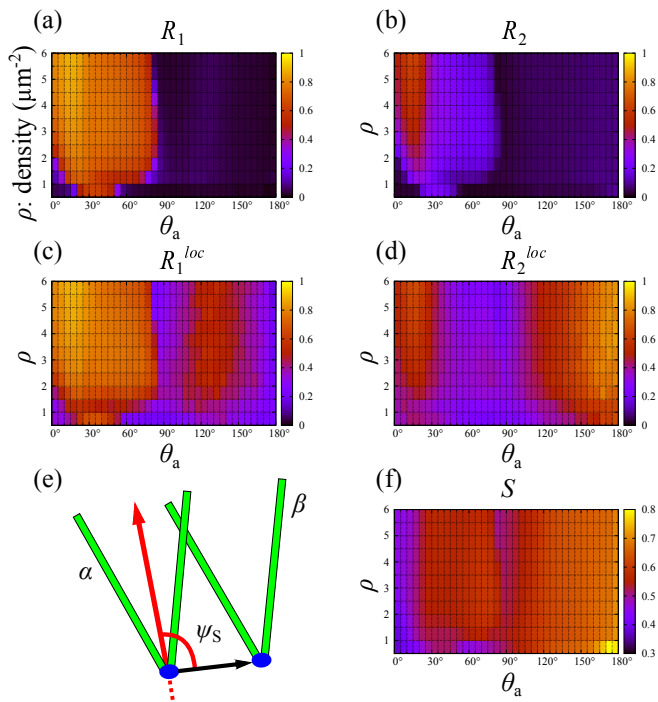}
\caption{\label{fig:order1}
(a)--(d) Color maps of $R_1$, $R_2$, $R_1^{loc}$, and $R_2^{loc}$ on $\theta_a$ and $N$.
(e) Definition of $\psi_S$. (f) Color map of $S = \langle \sin^2 \psi_S\rangle$.
The parameter values were calculated by the time average between $900~{\rm s} < t < 1000~{\rm s}$.
}
\end{figure}

%%%%%%%% Collective Motios of AFCs  %%%%%%%%%%%%%%
\section{Collective motions of AFCs}\label{sec:pattern}
Several types of collective behaviors were observed depending on the shape angle $\theta_a$ and density $\rho = N/L^2 ~{\rm \mu m^{-2}}$. In Fig.~\ref{fig:snap1}, examples of observed dynamics are shown in snapshots at the indicated parameters values.
In the figures, each AFC is colored according to its direction $\psi_{i}~(i = 1,2, \cdots, N)$ defined as the bisecting direction of the two filaments [Fig.~\ref{fig:snap1}(a) and (b)].
As explained below, we measured several quantities that characterize the AFCs' collective behaviors and classified the dynamics into four phases:
globally ferromagnetic [Fig.~\ref{fig:snap1}(c)], moving smectic [Fig.~\ref{fig:snap1}(d)], disorder [Fig.~\ref{fig:snap1}(e)], and locally nematic [Fig.~\ref{fig:snap1}(f)] patterns.

\subsection{Characterization of AFC dynamics}
The most significant quantity for characterization of AFC
dynamics is the orientation order parameters. We measured ferromagnetic
(polar) and nematic (apolar) parameters, $R_1$ and $R_2$, defined
by $ R_1 \equiv \frac{1}{N} \Bigl| \sum_{i = 1}^N\exp{(\mathrm{i}\psi_i)} \Bigr| $ and $ R_2 \equiv \frac{1}{N}\Bigl| \sum_{i = 1}^N\exp{ (2\mathrm{i}\psi_i)} \Bigr|$, respectively.

In some parameter regions, the system does not exhibit global order
but rather shows strong directional alignment among neighboring AFCs
[Fig.~\ref{fig:snap1}(f)].
To detect such local structures, we
divided the system into regular $10 \times 10$ lattices with
  filament length size ($\ell \times \ell ~\rm{[\mu m}^{2}]$),
measured the ferromagnetic and nematic order
parameters in each region, and then averaged their values.
These local order parameters $R_1^{loc}$ and $R_2^{loc}$ are given as $ R_1^{loc} \equiv \frac{1}{100} \sum_{j = 1}^{100} \frac{1}{N}_j\Bigl| \sum_{i=1}^{N_j}\exp{(\mathrm{i}\psi_i)} \Bigr|$ and
$ R_2^{loc} \equiv \frac{1}{100}\sum_{j = 1}^{100} \frac{1}{N}_j\Bigl| \sum_{i=1}^{N_j}\exp{(2\mathrm{i}\psi_i)} \Bigr|$, respectively.
Here, $N_j$ indicates the number of AFCs whose anchor is in
the $j$-th lattice ($j = 1, 2, \cdots, 100$).
These quantities are shown on a color scale in Fig.~\ref{fig:order1}(a)--(d).

As we will see below, the system shows another spatial regularity in
the moving smectic phase where the density bands of the aligned AFCs
are observed [Fig.~\ref{fig:snap1}(c)].
To characterize the density bands in the AFCs' positions,
an additional order parameter, $S$, was
introduced as the average of $\sin^2 \psi_S$, where $\psi_S$ is the
angle defined for two contacting AFCs, as indicated in
Fig.~\ref{fig:order1}(e).
If two AFCs are aligned in parallel and are
positioned side by side, i.e., the line connecting their anchors is
perpendicular to their directions ($\psi_S \sim \pm 90^{\circ}$), $S$
becomes high.
The color map of $S$ is shown in Fig.~\ref{fig:order1}(f).

To confirm that our analyses are independent of the system size, we conducted numerical simulations with different system sizes.
The results are summarized in Appendix~\ref{sec:appA}.

\subsection{Dynamics of AFCs}
(a) \textit{ Globally ferromagnetic order}.~
For a small angle between the two filaments of an AFC ($0^\circ \leq \theta_a \leq 80^\circ$), $R_1$ becomes nonzero as the AFC's density increases, which indicates a globally ferromagnetic order [Fig.~\ref{fig:snap1}(c) and Fig.~\ref{fig:order1}(a)].
$R_1^{loc}$ also becomes nonzero at the same density; hence, once the order emerges, it grows globally, indicating a phase transition [Fig.~\ref{fig:order1}(c)].
Note that at $\theta_a = 0^\circ$, an AFC is equivalent to a single polar filament, for which ferromagnetic transition was reported in earlier studies \cite{Aranson2006, Ben-Naim2006}.

When the concentration of AFCs is high, directional order develops up to the system size, and defects such as vortices and asters are not observed.
It is possible that these defects would be observed in the lower density region near the transition point; however, it is practically difficult to detect these structures at the resolution required for this type of simulation.
Therefore, we did not consider vortices and asters further in the present study.

\begin{figure}[b]
\includegraphics[width=7.0cm]{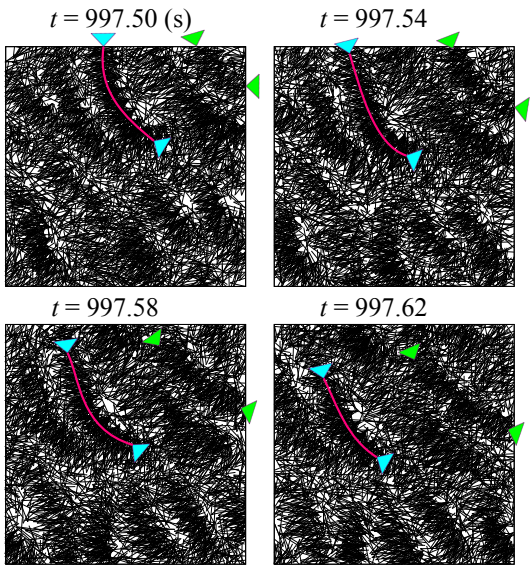}
\caption{\label{fig:snapDW}
Time evolution of the moving smectic structure ($\theta_a = 30^\circ$ and $\rho = 5~\rm{\mu m}^{-2}$).
Two lamellae are indicated by green and cyan arrowheads, respectively.
One lamella is traced with a pink line.
}
\end{figure}

(b) \textit{Moving smectic structure}.~
Inside the region of the ferromagnetic order phase, lamellar structures that consist of AFCs facing the same direction are observed (Fig.~\ref{fig:snap1}(d)).
The lamellae move through the system in one direction [Fig.~\ref{fig:snapDW}].
We call this dynamic pattern a `moving smectic' by analogy with the physics of liquid crystals.
Although moving band structures of density waves were reported in some actively propelling systems \cite{Vicsek1995,Chate2006,Peruani2011}, the smectic pattern found here is distinct, because the dominant propelling direction of each particle is nearly perpendicular to the band (See Sec.~\ref{sec:motion}).
As shown in Fig.~\ref{fig:order1}(f), the order parameter $S$ distinguishes this pattern from the globally ferromagnetic order where the positions of the AFCs exhibit no regularity.

(c) \textit{Disorder}.~
When the density of the AFCs is small, diffusion is dominant since the AFCs are sparse and thus interactions are rare.
There is neither orientational nor spatial order.
Even when the density of the AFCs is high, in the region $80^\circ \leq \theta_a \leq 120^\circ$, no sign of order in orientation, space, and time is detected [Fig.~\ref{fig:snap1}(e), Fig.~\ref{fig:order1}(f)].

(d) \textit{Locally nematic structure}.~
For larger $\theta_a$ up to $180^\circ$, $R_2^{loc}$ becomes nonzero, while $R_2$ remains zero regardless of the AFCs' density [Fig.~\ref{fig:order1}(b) and (d)].
The snapshot in Fig.~\ref{fig:snap1}(f) clearly indicates that neighboring AFCs are aligned and bundled, whereas orientational order is not maintained at a scale larger than the length of a few filaments.

\begin{figure}[b]
\includegraphics[width=8.6cm]{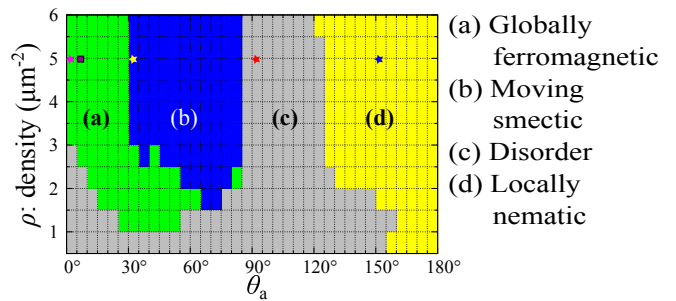}
\caption{\label{fig:PD} The phase diagram of the collective patterns of AFCs.
(a) Globally ferromagnetic, (b) moving smectic, (c) disorder, (d) locally nematic.
The precise criterion for assignment of phases in the numerical simulation is based on time-averaged order parameters as follows:
globally ferromagnetic: $R_1 > 0.5$ and $S < 0.55$;
moving smectic: $R_1 > 0.5$ and $S > 0.55$;
locally nematic: $R_1 < 0.5$ and $R_2^{loc} > 0.5$;
disorder: otherwise.
Stars in the figure indicate the parameters used for analyzing dynamics in the respective phases. The square indicates the parameters used in Fig.~\ref{fig:skew}.
}
\end{figure}

\subsection{\label{subsec:PD}Phase diagram}
A phase diagram of the observed dynamics is depicted in Fig.~\ref{fig:PD}.
For the high-density region $(\rho > 3 {\rm \mu m^{-2}})$, disorder/ferromagnetic transition is observed with small $\theta_a$, $0^\circ \leq \theta_a \leq80^\circ$.
A moving smectic structure appears in the higher density region of $30^\circ \leq \theta_a \leq 80^\circ$.
As $\theta_a$ increases, these two patterns vanish and the motions become disorder.
For large obtuse angles of $\theta_a$, locally nematic orders arise in $120^\circ \leq \theta_a \leq 180^\circ$.
The detailed characteristics of each pattern are summarized in Table~\ref{tab:table1}.
In the remainder of this paper, the density of AFCs is fixed at $\rho = 5~\mu m^{-2}$, where all the dynamics are observed by changing $\theta_a$.
The dynamics of globally ferromagnetic, moving smectic, disorder, and locally nematic patterns are analyzed by setting $\theta_a = 0^{\circ}, 30^{\circ}, 90^{\circ},$ and $150^{\circ}$, respectively (represented as stars in
Fig.~\ref{fig:PD}).

%%%%%%%%%%%%%%

\section{\label{sec:motion}Individual motion of AFCs}

\begin{figure}[b]
\includegraphics[width=8.4cm]{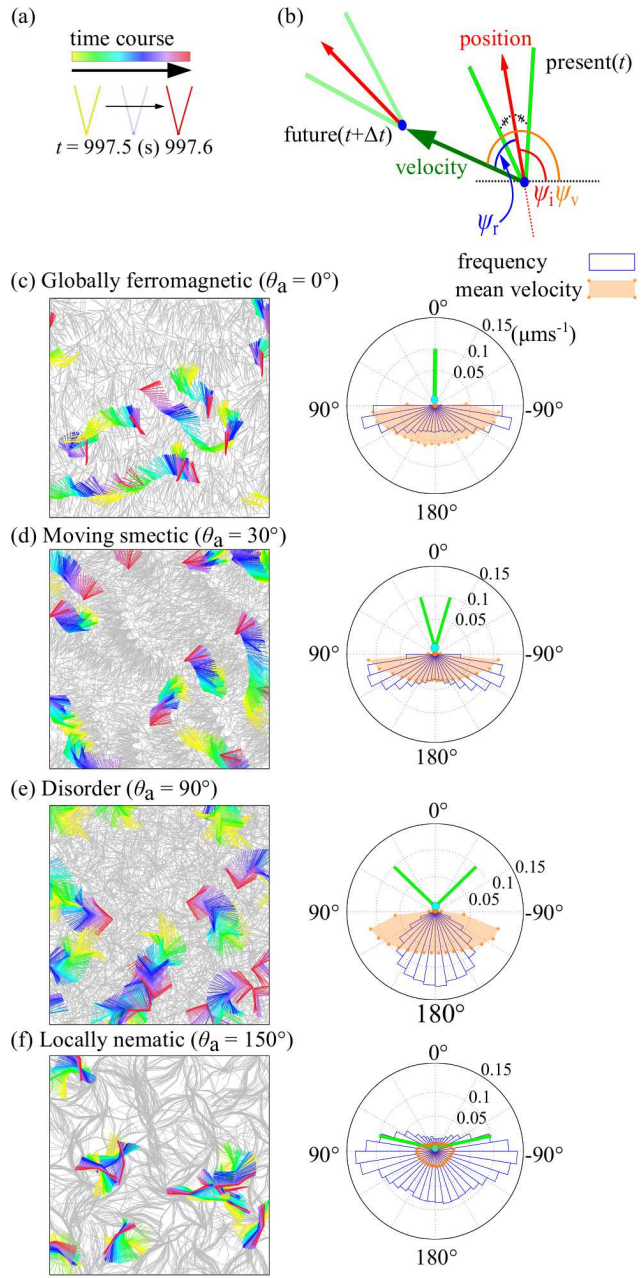}
\caption{\label{fig:orbit1}
Trajectories of individual AFCs in the collective motion.
(a) Color code of trajectories. AFCs are colored along the time course from yellow to blue, and finally red.
(b) The definition of $\psi_i$, $\psi_v$, and $\psi_r$.
(c)--(f) (left) Trajectories of individual AFCs in the indicated phase. Ten AFCs are colored according to (a); the others are colored gray.
(right) The frequency of $\psi_r$ (blue) and the average magnitude of velocity with respect to $\psi_r$ (orange).
}
\end{figure}

To further understand the collective dynamics of AFCs described in the preceding section, the individual motions of the AFCs were studied by tracking their trajectories.
Employing the color code indicated in Fig.~\ref{fig:orbit1}(a), the trajectories of $10$ AFCs are shown in each dynamic [Fig.~\ref{fig:orbit1}(c)--(f), left column].
Using the trajectories, we measured the propelling direction of each AFC, $\psi_v$, determined by the positional shift of the anchor between two time points $t$ and $t + \Delta t$.
The value $\Delta t$ is set to be $0.01 ~\rm{s}$, during which the density pattern seldom moves and the AFCs more or less maintain the same contacting pairs [Fig.~\ref{fig:snapDW}].
Then, the relative propelling direction of an AFC is defined by $\psi_r = \psi_v - \psi_i$ [Fig.~\ref{fig:orbit1}(b)].
The frequency and mean velocity of an AFC in a given direction $\psi_r$ are summarized in Fig.~\ref{fig:orbit1}(c)--(f) (right column).

(a) \textit{Globally ferromagnetic order}.~
In the parameter region of the globally ferromagnetic order, the motion of an AFC is translational without significant rotational motion [Fig.~\ref{fig:orbit1}(c)].
Statistical analysis showed that they barely moved in the same direction as that of the AFCs.
Most frequently, the AFCs move almost transversely to their direction ($\psi_r \simeq \pm 105^\circ$) where the mean velocity is fastest.
The less frequently observed AFCs reversed their direction of movement ($\psi_r \simeq 180^\circ$) at slower velocity.

(b) \textit{Moving smectic structure}.~
The direction and magnitude of the translational motion of AFCs are similar to those observed in the globally ferromagnetic phase [Fig.~\ref{fig:orbit1}(d)].
In the moving smectic phase, where a large proportion of the AFCs are aligned in parallel in the layered bands of the lamellar structure, the statistics indicated that the AFCs are moving along the density band, and the density band itself moves slowly in the direction opposite to that of the AFCs [Fig.~\ref{fig:snapDW}].
%These individual molecular motions contrast with those reported in other systems \cite{Head2013}.

\begin{figure}[b]
\includegraphics[width=8.6cm]{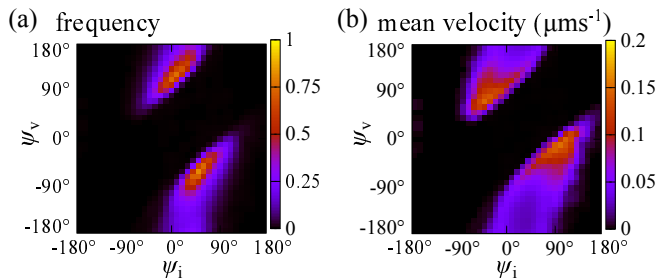}
\caption{\label{fig:posvel30}
Statistical analyses in the moving smectic phase.
(a) Joint distribution of $\psi_i$ and $\psi_v$.
The peak positions are at $(\psi_i,\psi_v) \simeq (60^\circ, -45^\circ),~(40^\circ, 145^\circ)$.
(b) Magnitude of the AFCs' velocity on $\psi_i$ and $\psi_v$.
The peak positions are at $(\psi_i,\psi_v)\simeq (110^\circ, 0^\circ), (-40^\circ,70^\circ)$.
The majority of the AFCs in the population move toward $\psi_v \simeq -45^\circ$ or $145^\circ$, which is insufficient for momentum to be conserved.
A few AFCs move faster toward $\psi_v \simeq 0^\circ$ or $70^\circ$, which compensates for the momentum conservation.
}
\end{figure}

These behaviors are apparently contradictory to the momentum conservation that must be satisfied in the present model, except for the noise effect.
For a more detailed analysis, the joint distribution and mean velocity on $\psi_i$ and $\psi_v$ were measured at a given time point and were compared.
The joint distribution shown in Fig.~\ref{fig:posvel30}(a) revealed two peaks at $(\psi_i,\psi_v) \simeq (60^\circ, -45^\circ), ~(40^\circ, 145^\circ)$, consistent with the high frequency of $\psi_r = \psi_v-\psi_i $ around $\pm 105^{\circ}$ [Fig.~\ref{fig:orbit1}(d)].
On the other hand, the magnitude of the mean velocity showed two peaks at $(\psi_i,\psi_v) \simeq (110^\circ, 0^\circ), (-40^\circ,70^\circ)$ [Fig.~\ref{fig:posvel30}(b)].
The discrepancy of the peak positions between frequency and velocity support the following scenario: although a large portion of the AFCs makes up the density bands that determine their collective behaviors, a smaller portion of the AFCs exhibit motion that is far from their collective average, which is much faster and in different directions, thereby compensating for the conservation law.
Indeed, AFCs that move fast and escape from a density band were observed in the simulation [Fig.~\ref{fig:snap1}(d)].
Such `division of labor' in AFC assembly is likely responsible for the appearance of a moving density band in the system with momentum conservation.

(c) \textit{Disorder}.~
In the disorder phase, the motion is still translational, although the predominant propelling direction now coincides with the direction opposite to that of the AFCs [Fig.~\ref{fig:orbit1}(e)].

(d) \textit{Locally nematic structure}.~
The motion in the locally nematic region is rotational rather than translational [Fig.~\ref{fig:orbit1}(f)].
Compared to the other phases, the velocity of translational motion is low.
Without significant translation of AFCs, the alignment process of AFCs occurs locally, resulting in the bundled structure of AFCs [Fig.~\ref{fig:snap1}(f)].

\section{\label{sec:two-body}Two-body interaction}

\begin{figure}[b]
\includegraphics[width=8.6cm]{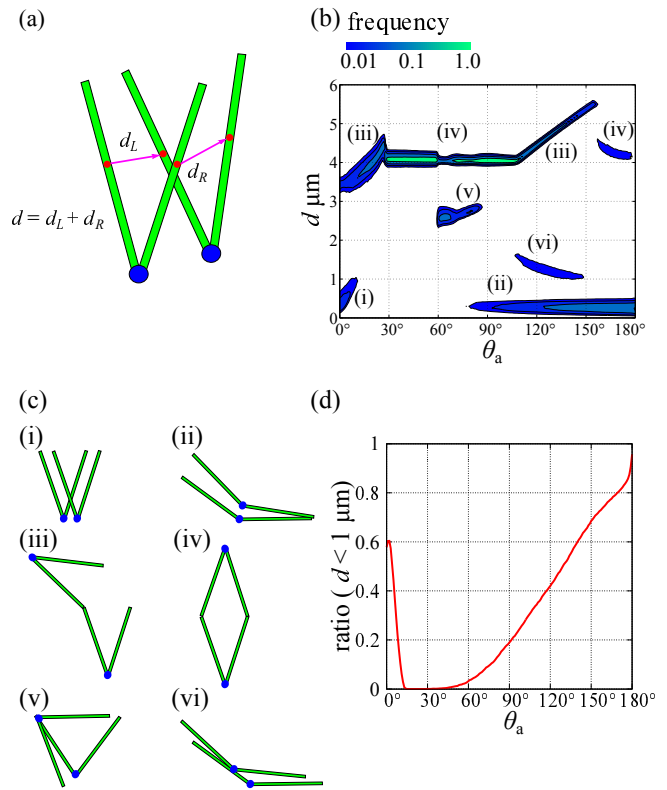}
\caption{\label{fig:2inter}
Analyses for a two-body interaction.
(a) The distance of the two AFCs is defined by $d = d_L + d_R $, where $d_L (d_R) $ is the distance between the centers of L-(R-)filaments.
(b) Distribution of $d$ after the simulation for each $\theta_a$. (i)--(vi) denote domains with characteristic configurations, as depicted in (c).
(d) The ratio that the final distance of two AFCs is smaller than $1 ~\rm{\mu m}$.
}
\end{figure}

Investigating the interaction between AFCs is crucial for understanding the emergence of various dynamics at different parameter regions.
In this section, we assess whether the completely overlapped state of two AFCs is sustained during the time evolution of the model.
Linear analysis is not useful in this case due to the condition that interaction works only when filaments are crossing; thus, the state is singular.
Instead, we numerically checked the stability of the overlapped state by measuring the final distance between the two initially overlapped AFCs.
The distance of the two AFCs is defined by the sum of distances between the centers of L-L and R-R filaments [$d = d_L + d_R ~\rm{\mu m}$, see Fig.~\ref{fig:2inter}(a)], by which $d = 0 ~\rm{ \mu m}$ corresponds to the overlapped state.
We measured the eventual distance of $d~\rm{\mu m}$ after simulating the two-AFCs system with a simulation time up to $t = 1000~s$. By iterating $10,000$ independent runs of the simulations starting from an almost overlapped but slightly perturbed state ($d < 0.01$),
the distribution of the distance was evaluated for a given value of $\theta_a$.
The results were robust to the choice of the simulation time as long as $t > 100~s$.

By calculating the distribution for each value of $\theta_a$,
we obtained the contour map of this distribution, as shown in Fig.~\ref{fig:2inter}.
Eventual states are classified into six domains, (i)--(vi), corresponding to the configurations of the two AFCs depicted in Fig.~\ref{fig:2inter}(c).
Configurations (i) and (ii) represent the almost overlapped and well-aligned states, where the distance $d$ is smaller than $1 ~\rm{ \mu m}$, which is equal to half of the filament length $\ell/2$.

In $\theta_a < 10^{\circ}$, two types of configurations, (i) and (iii), are possible;
however, the probability of type (iii) becomes higher as $\theta_a$ increases.
For $30^\circ \leq \theta_a \leq 80^\circ$, there are no domains with small $d$, indicating that the overlapped state is not maintained.
In particular, configuration (iv) dominates the probability up to $\theta_a < 120^\circ$.
For $\theta_a \geq 80^\circ$, an almost overlapped configuration (ii) appears, and the probability becomes higher as $\theta_a$ increases, whereas the probability in non-aligned configurations becomes smaller.
Within this region ($105^\circ \leq \theta_a \leq 150^\circ$), configuration (vi), in which two AFCs are loosely aligned, is also possible.

Figure \ref{fig:2inter}(d) (red lines) shows the ratio so that the eventual distance \REV{$d$} is smaller than $1 ~\rm{\mu m}$ [i.e., domains (i) and (ii)].
As $\theta_a$ increases from $\theta_a = 0^{\circ}$, configurations (iii) and (iv) become more dominant and
the ratio decreases to almost $0 ~\rm{\mu m}$ ($\theta_a > 10^\circ$).
The overlapped state is not stable where two AFCs move in opposite directions and are repulsive, which is responsible for the translational motions seen in the globally ferromagnetic, moving smectic, and disordered states.
The ratio increases with larger values of $\theta_a$. With the disappearance of configuration (iv) where $\theta_a > 120^\circ$, the overlapped state tends to be sustained as configuration (ii). This observation of interaction between two AFCs is critical for the local nematic pattern, in which translational motion is suppressed and the AFCs are well aligned.

\section{\label{sec:cross}Crossing of AFCs and filaments}

\begin{figure}[b]
\includegraphics[width=8.6cm]{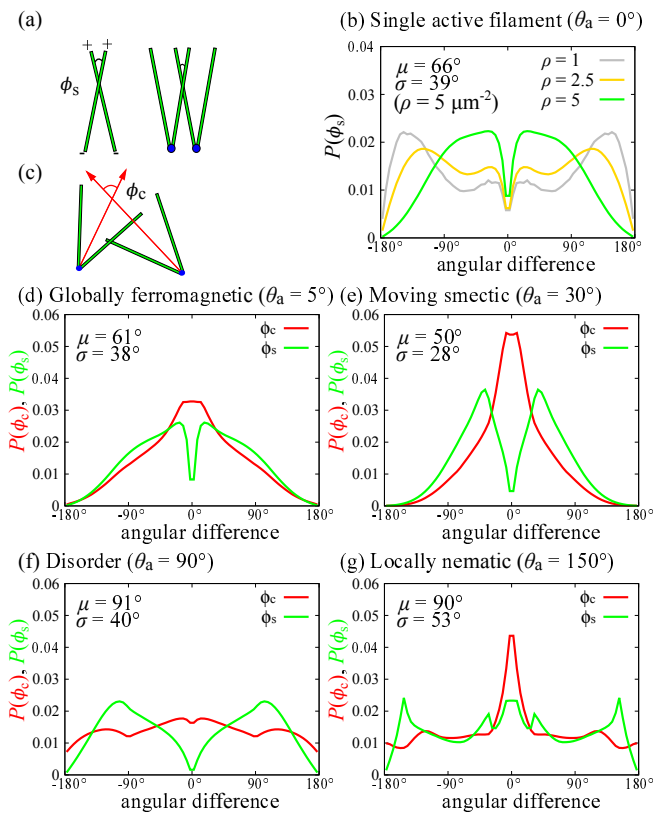}
\caption{ \label{fig:diffalignA}
Statistics describing how the AFCs and filaments cross each other.
(a) Definition of $\phi_s$, the angular difference between crossing filaments.
(b) Distribution of $\phi_s$ for single filaments on $\rho = 1, 2.5, 5 ~\rm{\mu m}^{-2}$.
(c) Definition of $\phi_c$, the angular difference between crossing AFCs.
(d)--(g) Distributions of $\phi_s$ and $\phi_c$ are denoted by green and red lines, respectively.
}
\end{figure}
\begin{figure}[b]
\includegraphics[width=8.6cm]{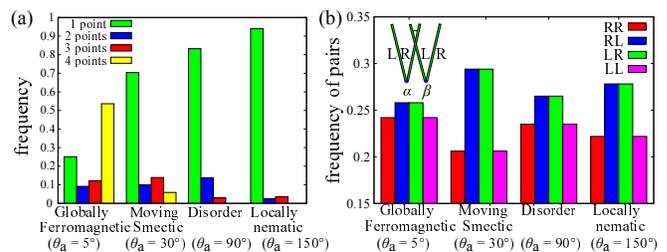}
\caption{\label{fig:skew}
Frequencies of the crossing types of AFCs.
(a) Frequencies of the number of crossing points between two contacting AFCs.
(b) Frequencies of crossing types. RL-crossing AFCs are shown in the figure.
}
\end{figure}

Having elucidated the interaction between two filaments, in this section we investigate how AFCs in a population cross each other to achieve characteristic patterns depending on the parameters.
In particular, because only the angle between crossing filaments determines the development of an AFC's position in Eq.~(\ref{eq:z1}), our main interest is the relative angle distributions among interacting AFCs.
Two kinds of angles are considered below.
The first is the angular difference between crossing filaments, $\phi_s$, as shown in Fig.~\ref{fig:diffalignA}(a).
The second is the angular difference between AFCs, $\phi_c$, measured between the directions of contacting AFCs [Fig.~\ref{fig:diffalignA}(c)].

Because an AFC has two filaments, complications can arise from the variety of crossing combinations for each filament.
Measurements of the number of crossing points between two contacting filaments revealed that in most cases, except when considering globally ferromagnetic dynamics, the filaments dominantly crossed at a single point [Fig.~\ref{fig:skew}(a)].
The abundance of four-point crossings in the globally ferromagnetic phase is easily explained by the small angle of $\theta_a$.
Therefore, we only consider one-point crossings and classify them into four types, namely RR, LR, RL, and LL.
The frequencies of these crossing types are shown in Fig.~\ref{fig:skew}(b).

(a) \textit{Single active filament system}.~
First, we investigate crossing in a single active filament system ($\theta_a = 0^\circ$) where $\phi_s$ and $\phi_c$ are identical.
Distributions of the angles are dependent on the density $\rho$ [Fig.~\ref{fig:diffalignA}(b)].
Despite the global order at $\rho = 5 ~\rm{\mu m}^{-2}$, the angular distribution $P(\phi_s)$ is not highest at $\phi_s = 0^\circ$, indicating that contacting filaments are not aligned in parallel but are instead tilted toward each other ($\phi_c \neq 0^\circ$);
the mean and standard deviation of the angular difference are calculated from the distribution of $|\phi_s|$ to be $\mu = 66^\circ$ and $\sigma = 39^\circ$, respectively.

(b) \textit{Global ferromagnetic order}.~
In a small angle of $\theta_a$ corresponding to the globally ferromagnetic phase, the distribution of $\phi_c$ becomes different from that of $\phi_s$ [Fig.~\ref{fig:diffalignA}(d)].
The angular distribution $P(\phi_c)$ now exhibits a peak at $\theta_c = 0^\circ$, which means that parallel alignment of two AFCs are enhanced more than for a single active filament owing to the change in molecular shape.
This enhancement in AFC alignment explains the increase in $R_1$ against $\theta_a$ in Fig.~\ref{fig:FSS}(a) (Appendix~\ref{sec:appA}) and the lower critical density for larger $\theta_a$ $(0^{\circ} < \theta_a < 80^{\circ})$ in Fig.~\ref{fig:PD}.

(c) \textit{Moving smectic structure}.~
The distribution of the angular difference between contacting filaments ($\phi_s$) is small ($\mu = 50^\circ$) and is sharply distributed ($\sigma = 28^\circ$).
With higher frequency of the LR and RL types of crossings compared to that of the other phases [Fig.~\ref{fig:skew}(b)], the distribution of $\phi_c$ shows a significantly high peak at $\phi_c = 0^\circ$, indicating a strong tendency of AFCs to be directed in parallel, which underlies the formation and maintenance of moving density bands.
Note that such parallel alignment of AFCs is not predicted from the two-body analysis (See Fig.~\ref{fig:2inter} in Sec.~\ref{sec:two-body}), and it results from many body interactions.

(d) \textit{Disorder}.~
The angle between crossing filaments, $\phi_s$, is most frequently found in the perpendicular direction $(\phi_s \sim 90^{\circ})$.
This distribution is almost identical to those observed for collections of single filaments with random orientation [Fig.~\ref{fig:diffalignA}(f)].
The frequency of more-than-two-points crossings is small [Fig.~\ref{fig:skew}(a)].
Consistent with a random orientation, the distributions of $\phi_c$ and crossing types [Fig.~\ref{fig:skew}(b)] are almost uniform.
These results indicate that orientational order does not exist, not even locally.

(e) \textit{Locally nematic structure}.~
In the locally nematic phase, local alignment processes make the AFCs
overlap and form a bundle structure.
In the bundle, the alignment is almost perfectly overlapped, as represented by a sharp peak in the
distribution of the angular differences of AFCs at $\phi_{c}=0^\circ$ [Fig.~\ref{fig:diffalignA}(h)].
The frequency of more-than-two-points crossings is negligibly small [Fig.~\ref{fig:skew}(c)].
In the distribution of angular differences in crossing filaments, several peaks appear [Fig.~\ref{fig:diffalignA}(h)].
These peaks can be explained by interactions of two AFCs, as shown in Fig.~\ref{fig:2inter}(b) and (c).
The peaks at $\phi_s = 0^\circ$ are due to LL/RR crossings corresponding to configuration (ii), whereas the peaks at $\phi_s = \pm \theta_a$ are due to LR/RL crossings corresponding to configuration (iii).
The other two peaks in the neighbors of $\phi_s = 0^\circ$ are produced by configuration (vi).

\begin{table*}
\caption{\label{tab:table1}Summary of emergent patterns (the light-anchor model).}
%\begin{ruledtabular}
\begin{tabular}{|c||c|c|c|c|}
  \hline
  Patterns & Globally & Moving & Disorder & Locally \\
  &ferromagnetic & smectic &  & nematic \\ \hline \hline
 $\theta_a$ & $0^\circ-80^\circ$ & $30^\circ-80^\circ$ & $80^\circ-120^\circ$ & $120^\circ-180^\circ$\\ \hline
 Density transition & $\bigcirc$ & $\bigcirc$ & $\times$ & $\times$ \\ \hline
 Characteristic & Translational & Translational & Translational & Rotational \\
 movement & perpendicularly & perpendicularly & parallelly & \\
  & to AFC & to AFC & to AFC &  \\ \hline
 Alignment & tilted & parallel & None & Strong \\
 b/w crossing AFC's & & & & 2-times symmetry \\ \hline
  Alignment & Isotropic & Perpendicular & None & None  \\
 propagation &  & to AFC & & \\ \hline
\end{tabular}
%\end{ruledtabular}
\end{table*}

\section{\label{sec:disc}Discussion}

In the majority of previous studies, objects with simpler shapes were assessed, such as material points, spheres, rods, or ellipsoids, whether stiff or flexible \cite{Ramaswamy2010,Marchetti2013}.
Shapes that are more complex, however, can play important roles in biological systems.
For example, the self-organization of microtubules $\sim \! \! 10~\rm{nm}$ beneath the apical membrane of multi-ciliated cells is associated with a regular array of cilia \cite{Kunimoto2012,Werner2011}.
In the present study, with the aim of understanding the dynamics of the molecular complexes of active filaments, we performed two-dimensional simulations of interacting AFCs defined as connected pairs of biofilaments, whose shapes were characterized by the angle between the two filaments.

The AFCs exhibited several types of collective patterns with global and local orders.
These patterns can be understood by tracing the trajectories of AFCs and investigating the alignment between crossing pairs of AFCs, as described in Sec.~\ref{sec:motion} -- \ref{sec:cross}.
Analysis of the two-body interaction in Sec.~\ref{sec:two-body} revealed that the overlapped state of AFCs was maintained where $\theta_a$ was large, which was responsible for the appearance of the local nematic order. In contrast, instability of the overlapped state indicates repulsion between the two AFCs, which is required for translational motion.
Translational motion is necessary for establishing the global order, because the phase transition of continuous order parameters in a two-dimensional system is possible only when a long-range interaction is present \cite{Mermin1966}.
As discussed in Sec.~\ref{sec:motion} and Sec.~\ref{sec:cross}, ferromagnetic orders can actually grow over a range that is much longer than the filament length if the AFCs move translationally and the alignment process can propagate.

Among the observed patterns, the moving smectic pattern is of particular interest.
In contrast to earlier studies that reported a moving density band \cite{Vicsek1995,Chate2006,Peruani2011}, in our system, the total momentum is conserved when the noise is eliminated.
The moving density band emerges even in the no-noise limit condition, which apparently contradicts with momentum conservation.
As described in Sec.~\ref{sec:motion}, a small population of AFCs that move fast in an atypical direction can compensate for the momentum, which leads to the appearance of a moving smectic.
In the present system, a complex molecular shape is responsible for this mechanism to arise.

The present study ignored some potentially important interactions for modeling experimental situations.
For example, it is possible that the length of filaments and the shape of AFCs are not uniform and instead change during the time evolution.
In particular, active regulation of filament length can be important for establishing ordered patterns, as exemplified by microtubule bundles in plant cells.
Furthermore, excluding the volume effect can result in alignment of the filaments and appearance of nematic order \cite{Liverpool2003,Liverpool2005,Ahmadi2005,Liverpool2006,Ahmadi2006}.
All of these processes can contribute to the organization of biofilament networks in actual systems.
In further research, the present analyses for AFCs could be easily extended, and this would likely provide useful insights for understanding collective patterns in more complex scenarios, including the aforementioned processes. Employing continuum description \cite{Tanase2004,Adhyapak2013,Chen2013,Menzel2014} will clarify the mechanism by which these patterns emerge as well as their stabilities.
The rapid developments in bio-nano engineering will also enable a comparison between experimental and simulated systems \cite{Dinu2009}.

\begin{acknowledgments}
The authors are grateful to N. Saito, K. Kaneko, S. Tsukita, and K. Oiwa for useful discussion. This work was supported by JST CREST `Creation of Fundamental Technologies for Understanding and Control of Biosystem Dynamics' and JSPS Grant-in-Aid for Scientific Research on Innovative Areas (25103008).
\end{acknowledgments}

\appendix

\section{\label{sec:appA}System size dependence}
\begin{figure}[b]
\includegraphics[width=6.0cm]{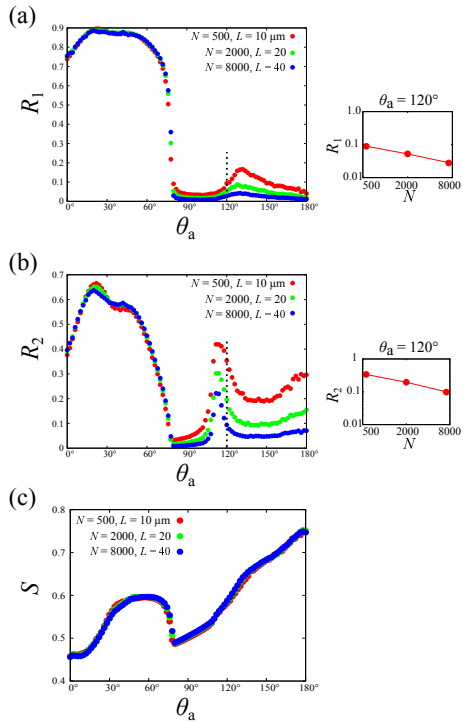}
\caption{\label{fig:FSS}
System size dependence of the dynamics.
$N$ and $L$ are varied by fixing the density $\rho = 5.0 ~\rm{\mu m}^{-2}$, as $(N,L) = (500,10 ~\rm{\mu m})$, $(2000,20)$, and $(8000,40)$.
(a)--(c) $R_1$, $R_2$, and $S$ are plotted against $\theta_a$ for different system sizes.
Right figures in (a) and (b) show the system size dependency of $R_1$ and $R_2$ at $\theta_a = 120^\circ$ (broken lines).
}
\end{figure}

To check the system size dependence of the dynamics observed in the main text, we conducted numerical simulations with different system sizes.
We fixed the density as $\rho = 5 ~\mu m^{-2}$ and changed the system size as $L = 10, 20,$ and $40 ~\mu m$. The results are shown in Fig.~\ref{fig:FSS} (a)--(c).
$R_1$ and $R_2$ on acute $\theta_a$ are independent of the system size, and finite size scaling analysis supports that the emergence of ferromagnetic order is a genuine phase transition, consistent with observation in rod-like filament dynamics \cite{Liverpool2006}.

On the other hand, $R_1$ and $R_2$ on obtuse $\theta_a$ decrease as the system size increases, suggesting that the globally nematic pattern is at most a quasi-long range order akin to an XY-model \cite{Chate2006,Peruani2011}.
$S$ is independent of the system size.

% The \nocite command causes all entries in a bibliography to be printed out
% whether or not they are actually referenced in the text. This is appropriate
% for the sample file to show the different styles of references, but authors
% most likely will not want to use it.
%\nocite{*}
%\bibliography{apssamp}% Produces the bibliography via BibTeX.
\bibliography{library_AFC1}

%
% ****** End of file apssamp.tex ******

\end{document}